\documentclass[%
 reprint,
superscriptaddress,
 amsmath,amssymb,
 aps,
]{revtex4-2}

\usepackage{physics} 
\usepackage{graphicx}
\usepackage{dcolumn}
\usepackage{bm}
\usepackage{color,sidecap,colortbl}
\usepackage[dvipsnames]{xcolor}
\newcommand{\kbar}{\mathchar'26\mkern-9mu k}
\usepackage{upgreek}
\usepackage{xcolor}

\begin{document}

\title{Observation of the Quantum Boomerang Effect}

\author{Roshan Sajjad}
\author{Jeremy L. Tanlimco}
\author{Hector Mas}
\author{Alec Cao}
\author{Eber Nolasco-Martinez}
\author{Ethan Q.\ Simmons}
\affiliation{Department of Physics, University of California, Santa Barbara, California 93106, USA}

\author{Fl{\'a}vio L. N. Santos}
\affiliation{Departamento de Física Teórica e Experimental, Universidade Federal do Rio Grande do Norte, 59072-970 Natal, Rio Grande do Norte, Brazil}

\author{Patrizia Vignolo}
\affiliation{Universit{\'e} C{\^o}te d'Azur, CNRS, Institut de Physique de Nice, 
06560 Valbonne, France}

\author{Tommaso Macr{\`i}}
\affiliation{Departamento de Física Teórica e Experimental, Universidade Federal do Rio Grande do Norte, 59072-970 Natal, Rio Grande do Norte, Brazil}

\author{David M.\ Weld}
\email{weld@ucsb.edu}
\affiliation{Department of Physics, University of California, Santa Barbara, California 93106, USA}
\begin{abstract}
A particle in an Anderson-localized system, if launched in any direction, should on average return to its starting point and stay there. Despite the central role played by Anderson localization in the modern understanding of condensed matter, this ``quantum boomerang'' effect, an essential feature of the localized state, was only recently theoretically predicted. We report the experimental observation of the quantum boomerang effect. Using a degenerate gas and a phase-shifted pair of optical lattices, we not only confirm the predicted dependence of the boomerang effect on Floquet gauge, but also elucidate the crucial role of initial state symmetries.  Highlighting the key role of localization, we observe that as stochastic kicking destroys dynamical localization, the quantum boomerang effect also disappears. Measured dynamics are in agreement with numerical models and with predictions of an analytical theory we present which clarifies the connection between time-reversal symmetry and boomerang dynamics. These results showcase a unique experimental probe of the underlying quantum nature of Anderson localized matter.
\end{abstract}

\maketitle

Anderson localization is a quintessential effect of disorder in a variety of physical systems~\cite{OGAnderson,fishman_dynamicallocalization,FishmanPRA1984,Haake2010,Casati1979,ALoflight,ALofsound,Billy2008,inguscio-andersonloc,ChabeAndersonTransition2008,DelandeAnderson2D,electronicd3DALexpt}. Disordered lattice models enjoy a robust theoretical proof of localization and remain a hotbed for discovery even today. While transport in disordered media has been studied for more than 50 years, a fundamental \emph{dynamical} feature of Anderson-localized matter known as the ``quantum boomerang'' effect was only recently theoretically predicted~\cite{PratPRA2019,JanarekPRA2020,TessieriPRA2021}. The central prediction of the quantum boomerang is that, for an appropriate average over disorder realizations, a wavepacket launched with any nonzero momentum will return to and localize at its initial position~\cite{PratPRA2019}. This surprising phenomenon can be understood as a consequence of both Anderson localization and time-reversal symmetry: the long-time density distribution for a wavepacket with an initial momentum $k_0$ (i) is time-independent because of localization and (ii) does not depend on the sign of $k_0$ because of time-reversal symmetry. Together, these features imply that at long times the wavepacket must settle at its initial position. The quantum boomerang effect is predicted to occur in a broad class of disordered or pseudo-disordered tight-binding models~\cite{TessieriPRA2021} including the quantum kicked rotor (QKR), a momentum-space realization of a disordered quasi-1D wire~\cite{GarreauReview2017,AltlandFieldTheory96} which provides an ideal platform for the study of the interplay between localization, chaos, interactions, and transport~\cite{RaizenQKR1995,HainautIdeal2019,HainautReturnToOrigin2017,HainautRatchet18,gaugefieldkickedrotor,WhiteRatchet2013,cao2021prethermal,toh2021observation,SummyRatchet2012,GueryOdelinCBSCFS,quasiperiodicqkr_3dmetaltoinsulator_exp2,victorMBDL,dallatorrepretherm}. In any context, the quantum boomerang effect contravenes classical predictions and serves as a sensitive diagnostic of the quantum nature of localization, with the potential, for example, to map mobility edges~\cite{PratPRA2019,DelandeMobilityEdge}. 

\begin{figure}[t!]
    \centering
    \renewcommand{\figurename}{Fig.}
    \includegraphics[scale = 1]{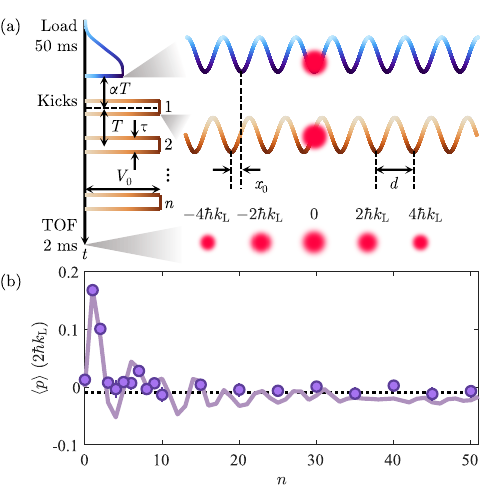}
    \caption{Observation of the quantum boomerang effect.
    \textbf{(a):} Experimental sequence. A BEC is loaded into one optical lattice and then kicked with a second, phase-shifted lattice. The momentum distribution is measured after free expansion. BEC size and timeline are not to scale.
    \textbf{(b):} Measured average momentum versus kick number, for parameters given in the text. The return to the origin demonstrates the quantum boomerang effect. Error bars show standard error of the mean on ten repeated measurements. Solid curve is the numerical prediction. Dotted line represents infinite-time average predicted by the Floquet diagonal ensemble (Eq. \ref{eq:FDE}).
    }
    \label{fig:setupBoomerang}
\end{figure}

We report the experimental observation and characterization of the quantum boomerang effect. Using a degenerate gas and a pair of phase-shifted optical lattices to create a kicked rotor with tunable initial conditions (Fig.~\ref{fig:setupBoomerang}), we observe the boomerang's distinctive return dynamics. We experimentally characterize the dependence of the dynamics on symmetry properties, Floquet gauge, and initial state, and confirm that results agree with an analytical theory we develop in the appendices. Finally, we measure the controlled destruction of the boomerang effect by stochastic kicking. The atom-optics kicked-rotor realization of the quantum boomerang offers three key advantages: (i) finite-width quasimomentum distributions naturally achieve single-shot disorder averaging, (ii) working in momentum space rather than position space enables simple observation of boomerang dynamics with single-site resolution, and (iii) symmetry properties can be straightforwardly tuned by adjusting properties of the quench~\cite{Sadgrove2007,Dana2008,Delvecchio2020}.  

\begin{figure*}[t!]
    \centering
    \renewcommand{\figurename}{Fig.}
    \includegraphics[scale = 1]{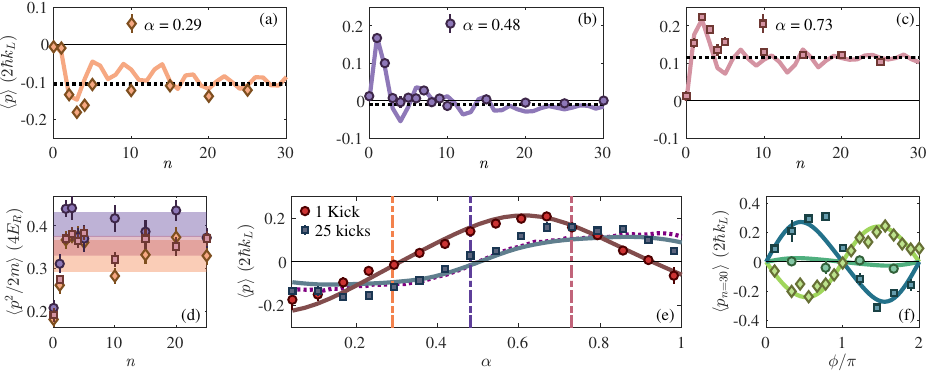}
    \caption{Characterization of the quantum boomerang effect. All error bars show standard error of the mean on repeated measurements. \textbf{(a-c):} Average momentum vs.\ kick number for $V_0=41\,E_\mathrm{R}$  and $T=8\,\upmu$s, at $\alpha=0.29$, 0.48, and 0.73 respectively. Solid curves show numerical predictions. Dotted lines show infinite-time average momenta predicted by the Floquet diagonal ensemble (Eq.~\ref{eq:FDE}). $\phi = 0.9 \pi/2$ for all experimental runs. 
    \textbf{(d):} Measured energy vs.\ kick number for all three values of $\alpha$. Shaded areas indicate long-term energy by averaging data points between 10 and 25 kicks.
    \textbf{(e):} Measured average momentum after 1 kick (red circles) and 25 kicks (blue squares) vs.\ $\alpha$ for the same experimental parameters. Same-color solid lines show numerical predictions. Dashed vertical lines indicate $\alpha$ in (a-c). Dotted purple line shows prediction of the Floquet diagonal ensemble~\cite{PratPRA2019}. \textbf{(f):} Average momentum after 30 kicks vs.\ $\phi$ for $V_0=70\,E_\mathrm{R}$, $T=7\,\upmu$s, and $\alpha = 0.15$  (diamonds), 0.48 (circles), and 1.05 (squares). 
        }
    \label{fig:boomerang}
\end{figure*}

We begin each experiment by preparing a Bose-Einstein condensate (BEC) of $10^5$ $^7$Li atoms and then tuning interatomic interactions to zero with a Feshbach resonance. A typical experimental sequence from this point on is depicted in Fig.~\ref{fig:setupBoomerang}a. The atoms are exposed in turn to two coaxial optical lattices with lattice constant $d=532\,\mathrm{nm}$, wave vector $k_\mathrm{L}=\pi/d$, recoil energy $E_\mathrm{R}=\hbar^2k_\mathrm{L}^2/2M$, and tunable relative offset $x_0$. $M$ is the atomic mass. To prepare an initial wavefunction which is well-localized in position space within each lattice site, we adiabatically load the BEC into the first optical lattice over $50\,\mathrm{ms}$, reaching a final depth of $7\,E_\mathrm{R}$. At $t=0$, we suddenly turn off the first lattice and quench into a Hamiltonian created by repeated kicking with the second, spatially-shifted lattice: 
\begin{equation}
\label{eq:ourHamiltonian}
H_\alpha=\frac{p^2}{2M}+\frac{V_0}{2}\cos\left(2k_\mathrm{L}x\right)\sum_nf_\tau(t-nT-\alpha T).
\end{equation}
As described by Eq.~\ref{eq:ourHamiltonian}, the second, phase-shifted lattice is pulsed $n$ times to depth $V_0$ with pulse shape $f_\tau$, effective pulse width $\tau=500\,\mathrm{ns}$, and period $T$; the first pulse peaks a delay time $t=\alpha T$ after the quench. Having measured the finite time of this quench to be approximately $190\,\mathrm{ns}$, we define $t=0$ halfway through the quench and recognize a systematic $\pm0.01$ uncertainty in the reported values of $\alpha$. Because of the adiabatic loading into the first lattice and the relative phase shift, the atomic ensemble starts out centered at a nonzero position given by $x_0$. This is analogous to the situation in a position-space quantum boomerang, where the initial state is centered at nonzero momentum.

To make concrete the connection to the kicked rotor we introduce dimensionless variables $P\equiv 2k_\mathrm{L}Tp/M$, $Q\equiv 2k_\mathrm{L}x$, and $\mathcal{T}\equiv t/T$ as well as dimensionless parameters $\kbar\equiv 8E_\mathrm{R}T/\hbar$, an effective Planck's constant, and $K\equiv\kbar V_0\tau/2\hbar$, encoding kicking strength. Substituting these into Eq.~\ref{eq:ourHamiltonian}, rescaling the Hamiltonian $\mathcal{H}_\alpha\equiv (4k_\mathrm{L}^2T^2/M) H_\alpha$, and taking the $\delta$-function limit of $f_\tau/\tau$, we recover the canonical QKR Hamiltonian
\mbox{$\mathcal{H}_{\alpha}=\frac{P^2}{2}+{K}\cos Q \sum_n\delta(\mathcal{T}-n-\alpha)$}.
This gives rise to a single-particle 1-cycle Floquet operator
\mbox{$U_\alpha=e^{-i(1-\alpha) P^2/2\kbar}e^{-iK\cos Q /\kbar}e^{-i\alpha P^2/2\kbar}$}.

Figure \ref{fig:setupBoomerang}b illustrates the main result of this work: we experimentally observe the departure from and return to zero average momentum which is the key signature of the quantum boomerang effect~\cite{PratPRA2019}. Here $V_0=41\,E_\mathrm{R}$, $T=8\,\upmu$s, $\phi\equiv 2k_\mathrm{L}x_0=0.9\pi/2$, and $\alpha=0.48$, which correspond to $\kbar=10$, $K=9$.  While the first kick imparts momentum asymmetry due to the lattice phase offset, the system returns to zero average momentum at long times. The experimental result agrees well with  fit-parameter-free time-dependent Schr{\"o}dinger equation (TDSE) simulations of the finite-pulse width Hamiltonian (\ref{eq:ourHamiltonian}) using a split-step method described in appendix \ref{sec:numerics}.

This demonstration of the quantum boomerang effect enables an in-depth experimental characterization of the circumstances under which it can occur. Our measurements reveal a rich interplay among symmetries and quench dynamics which is  well described by the analytic theory we present in the appendices. Specifically, it is the relation between the symmetry properties of the initial state and the Floquet gauge $\alpha$ defined by the $t=0$ quench which dictates the presence or absence of the quantum boomerang effect in the quantum kicked rotor. This interplay offers a contrast to static disordered systems, in which initial state symmetries are also important but which lack a Floquet gauge degree of freedom. We note for clarity that the QKR Hamiltonian itself is of course time-reversal and parity symmetric independent of Floquet gauge and spatial phase choices, so long as one references appropriate symmetry axes.  For the remainder of the paper, we will solely refer to time-reversal and parity symmetry with respect to inversions about $t=0$ and $x=0$ unless otherwise stated. $\mathcal{H}_{\alpha}$ is then clearly parity symmetric, and only when $\alpha=1/2$ does the evolution also become time-reversal invariant. As we show in Appendix \ref{sec:alpha=1/2}, the initial Bloch state ensemble we experimentally prepare satisfies the exact properties necessary to observe a boomerang effect under these conditions: it is (i) parity asymmetric to allow for nontrivial early-time momentum dynamics and (ii) time-reversal symmetric ($\psi^*(x) = \psi(x)$) to ensure a complete return to 0 average momentum at late-times. This interplay of symmetries arising from the QKR quench protocol enables the quantum boomerang effect to be switched on or off by adjusting the Floquet gauge.

Fig.~\ref{fig:boomerang} demonstrates this dependence on Floquet gauge with a detailed characterization of the average momentum evolution for various $\alpha$. Measurements at $\alpha$ values of 0.29, 0.48, and 0.73 (Fig.~\ref{fig:boomerang}a-c) exhibit fundamental differences in the dynamical behavior. As discussed above, only near $\alpha = 1/2$ do we observe the quantum boomerang effect: a sharp initial increase in $\langle p\rangle$ followed by a decline back to the initial value of 0. Resolution of the precise boomerang condition is limited experimentally by timing noise on the location of the first pulse and the finite quench time of the acousto-optic modulators.  In contrast, evolution under the other two Floquet gauges converges to nonzero momentum values at long times, indicating the absence of the full quantum boomerang effect. Each data set agrees well with fit-parameter-free numerical predictions, and all late-time momenta match the analytical predictions of the Floquet diagonal ensemble (Eq.~\ref{eq:FDE}).

Fig.~\ref{fig:boomerang}d shows that the observed energy evolution is qualitatively consistent for all three values of $\alpha$, with each ensemble exhibiting dynamical localization  as expected for a non-interacting quantum kicked rotor~\cite{Shepelyansky1986,RaizenQKR1995,cao2021prethermal}. We note that the energy for $\alpha=0.48$ appears to saturate at a slightly higher value than the other gauges, a result consistent with the predictions of Ref.~\cite{TessieriPRA2021}.

\begin{figure}[t!]
    \centering
    \renewcommand{\figurename}{Fig.}
    \includegraphics[scale = 1]{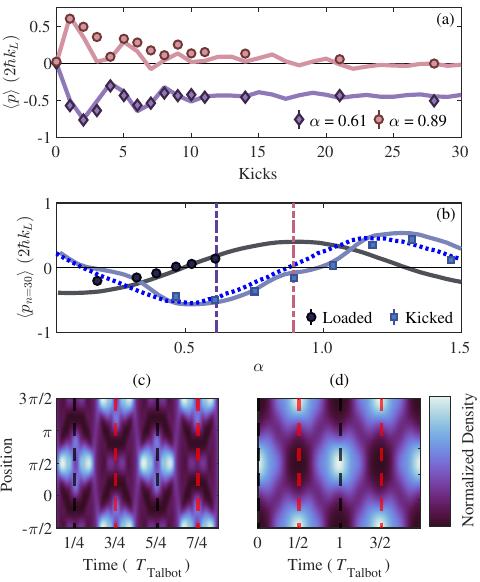}
    \caption{Initial state dependence of the quantum boomerang effect. All error bars show standard error of the mean on repeated measurements. 
    \textbf{(a):} Average momentum versus kick number for an initial state consisting of a BEC kicked once by a  $60\,E_\mathrm{R}$ lattice. Pink circles correspond to $\alpha = 0.89$ and purple diamonds to $\alpha = 0.61$ for $T=7\,\upmu\mathrm{s}$.  Solid lines are predictions of TDSE numerics.
    \textbf{(b):} Average momentum after 30 kicks versus $\alpha$, for an initial condition set by adiabatic loading of a BEC into a $7\,E_\mathrm{R}$ lattice (grey circles) and an initial condition set by kicking a BEC once with a $60\,E_\mathrm{R}$ lattice (blue squares). Dotted blue line shows infinite-time average momenta predicted by the Floquet diagonal ensemble (Eq.~\ref{eq:FDE}). Dashed vertical lines correspond to $\alpha$ values in (a). \textbf{(c-d):} Calculated density evolution over two Talbot times of a BEC kicked by a $60\,E_\mathrm{R}$ lattice (left) and a a BEC adiabatically loaded into a $7\,E_\mathrm{R}$ lattice which is then quenched off (right). Dashed vertical lines represent time-reversal symmetry axes.}
    \label{fig:kickprep}
\end{figure}

A key feature of quantum boomerang dynamics which can be probed with a kicked rotor is the contrast between short- and long-time evolution. Fig.~\ref{fig:boomerang}e maps out the Floquet gauge dependence of momentum in these two regimes. 
Since the energy saturates after only five kicks, we take 25 kicks as a reasonable proxy for the infinite-time limit predicted by relaxation to the Floquet diagonal ensemble (see Eq.~\ref{eq:FDE}). 
The initial momentum asymmetry imparted to the rotor after one kick arises purely from the Talbot effect, a periodic re-imaging of the wavefunction in time under free-evolution as a result of the spatially periodic lattice exciting only quantized momentum modes; this Talbot effect is intimately related to the extensively explored quantum resonance phenomena in the QKR~\cite{GarreauTalbotPRA2008,TalbotEffect1,TalbotEffect2,summy-resonance}. The Talbot oscillation for a Bloch state is illustrated in Fig.~\ref{fig:kickprep}d, where for our experiment, $T_{\mathrm{Talbot}} = \pi \hbar/2 E_{\mathrm{R}} \approx 9.95\,\upmu$s.

In contrast, the long-time behavior is governed by localization physics and symmetries surrounding the quench. One feature of both experimental and theoretical results in Fig.~\ref{fig:boomerang}e is that the long-time momentum should be an odd function about $\alpha=1/2$, a result we prove in appendix \ref{sec:palpha}. This result is connected to the resemblance between forward time-evolution in the gauge $\alpha$ (propagation with $U_{\alpha}$) and backward time-evolution in the gauge $1-\alpha$ (propagation with $U_{1-\alpha}^{\dagger}$). The anti-symmetry about $\alpha=1/2$ is dictated by the time-reversal invariance of the initial Bloch state ensemble. We note that because this property arises due to similarities between forward evolution in one gauge and backward evolution in the other, it only manifests as an (anti-)symmetry in the forward evolution of both gauges in the long-time limit (see Eq.~\ref{eq:p_inf=p_-inf}); at short times, no symmetry is generically expected, as evidenced by the 1-kick data.

The odd parity of the late-time momentum about $\alpha=0.5$ is further reflected in Fig.~\ref{fig:boomerang}f, which shows the late-time momentum as a function of initial wavepacket position for three different values of $\alpha$.  Here a shift in the relative lattice phase is exactly analogous to variation in the initial launch velocity of a position space boomerang. We observe that regardless of initial position, $\alpha$ values near 0 and 1 result in opposite-sign, non-zero late-time momenta. The only exception occurs when the initial state recovers parity symmetry at $\phi=0$ or $\phi=\pi$ and the average momentum dynamics become trivial. Near $\alpha = 1/2$, the entire curve collapses toward 0. This indicates that, as mentioned above, it is time-reversal symmetry which plays the fundamental role in determining the \emph{return} of the boomerang. The initial position which controls the degree of parity asymmetry is instead more closely related to the \emph{amplitude} of the early-time boomerang dynamics.

The demonstrated control of symmetry properties by varying $\alpha$ opens up the possibility of realizing quantum boomerang dynamics starting from more exotic initial states. Fig.~\ref{fig:kickprep} explores the quench dynamics of an initial state with an essentially homogeneous distribution in position space, achieved by replacing the adiabatic lattice load with a single kick. The initial wavefunction, approximately $\psi(x) = e^{-i V \sin(2k_{\mathrm{L}}x)}$ with $V$ the effective kick impulse, is neither time-reversal symmetric nor well-localized in position.  Intuitively one would not expect such a state to demonstrate quantum boomerang dynamics, as our understanding of the position-space quantum boomerang effect is based on a well-defined initial velocity. Despite these expectations, the data in Fig.~\ref{fig:kickprep}a reveal a clear quantum boomerang effect, though at a Floquet gauge $\alpha =0.86$, very different from the value of 0.5 expected for a time-reversal symmetric state. This can be understood by observing that free evolution maps this initial state into a time-reversal but not parity invariant wavefunction after a time $T_{\mathrm{Talbot}}/4$, as illustrated numerically in Fig.~\ref{fig:kickprep}c and proven analytically in Appendix \ref{sec:tablot/4_kickedstate}. Thus, one expects to observe the boomerang effect at $\alpha= 1/2+T_{\mathrm{Talbot}}/4T\simeq 0.855$ for the kicked initial state, in good agreement with numerical and experimental results in Fig.~\ref{fig:kickprep}. This finding demonstrates that even for quenched states completely unlike traditional boomerang starting points, the Floquet gauge freedom can restore the boomerang effect in certain cases where the Talbot effect defines a new time-inversion symmetric point. 

We note that while the kicked state is not initially time-reversal symmetric, it is invariant under simultaneous parity-time inversion ($\psi^*(-x) = \psi(x)$). As we show in Appendix \ref{sec:palpha}, this different symmetry condition manifests in the late-time momentum exhibiting even rather than odd symmetry about $\alpha=1/2$. These different Floquet gauge properties for the two initial state symmetries we consider are demonstrated and contrasted in Fig.~\ref{fig:kickprep}b, which clearly distinguishes the different parity about $\alpha=0.5$ (see also data for the adiabatically loaded initial condition plotted for a broader range of $\alpha$ in Fig.~2e). We note that the data are not perfectly symmetric on the appropriate intervals, as demonstrated by comparison to the Floquet diagonal ensemble predictions. We attribute this to the fact that the symmetry is only established in the infinite-time limit, with deviations present at any finite time. Nevertheless, such deviations from the long-time limit are small, and our finite-time observations clearly reflect the idealized theoretical predictions.  For the kicked state data shown, the symmetry is most apparent when extending to cases where $\alpha$ is slightly greater than 1~\footnote{
The experimental delay-time definition and Floquet gauge definition of $\alpha$ is only consistent on $\left[0,1\right]$ . For $\alpha>1$ data and simulations shown, we define $\alpha$ solely by the delay-time definition. To make the Floquet gauge definition consistent with this, one should redefine the quenched initial state as $e^{-i \left\lfloor \alpha \right\rfloor P^2/2\kbar} \ket{\psi_0}$, with $\ket{\psi_0}$ the original quenched state for $\abs{\alpha} \leq 1$ cases, $\left\lfloor \alpha \right\rfloor$ denoting the integer part of $\alpha$, and taking the Floquet gauge to be $\alpha \bmod 1$.
}.

\begin{figure}[tb]
    \centering
    \renewcommand{\figurename}{Fig.}
    \includegraphics[scale = 1]{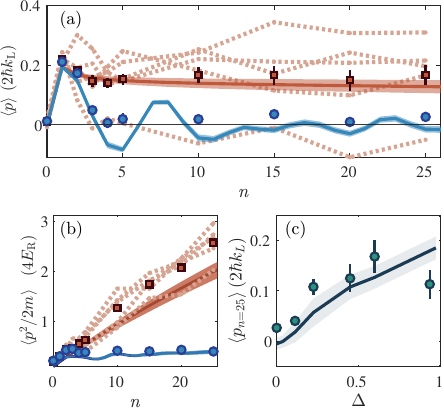}
    \caption{Destruction of the quantum boomerang effect by stochastic kicking. All error bars correspond to the standard error of the mean for all realizations of randomness.
    \textbf{(a):} Average momentum measured after $n$ kicks for temporal disorder strength $\Delta=0$ (blue circles) and $\Delta=0.6$ (red squares). For all data in this figure, $T=8\,\upmu$s, $\alpha=0.52$, $V_0=41\,E_\mathrm{R}$, and the initial condition was set by an adiabatic lattice load to $8\,E_\mathrm{R}$. Circles (squares) represent the mean of measurements with 5 separate kick disorder realizations for $\Delta=0$ ($\Delta=0.6$). Dashed lines show individual realizations. Shaded areas in (a) and (c) show numerically calculated $\langle p \rangle \pm \sigma_{\langle p \rangle}$ with 9000 disorder realizations and 500 initial quasimomenta. The shaded area in (b) includes an additional estimate of the energy accounting for the statistical error in the lattice depth of $V_0=41\pm2\,E_\mathrm{R}$ over 500 lattice depths and 500 realizations. 
    \textbf{(b):} Measured mean energy for $\Delta=0$ (blue circles) and $\Delta=0.6$ (red squares) over 10 points (blue) and 5 disorder realizations (red), illustrating the destruction of dynamical localization by random kicking.
    \textbf{(c):} Average momentum $\langle p \rangle$ after 25 kicks versus disorder strength $\Delta$. Dots are averages over 10 disorder realizations, except for $\Delta=0,0.6$ (5 realizations each). The solid line is numerically calculated $\langle p \rangle$ over 9000 disorder realizations, with shaded area showing a one-standard-deviation range.
    }
    \label{fig:theRandomness}
\end{figure}

The final set of results we describe (Fig.~\ref{fig:theRandomness}) probes the question: what happens to quantum boomerang dynamics as localization is destroyed?  In the QKR context, dynamical localization can be tunably disrupted by introducing temporal stochasticity into the kicking sequence~\cite{RaizenRandom2003,cao2021prethermal}. To implement this, we add a deviation $\delta_n$ to the delay time after the $n^\mathrm{th}$ kick, where $\delta_n$ is randomly selected from a uniform distribution in the range $\left[-W/2,W/2 \right]$. The parameter $\Delta=W/T$ quantifies the degree of randomness and can range from 0 to 1. Fig.~\ref{fig:theRandomness}a shows the time evolution of average momentum for $\Delta=0$ (no randomness) and $\Delta=0.6$, with the first kick fixed so that $\alpha=0.52$, and with the same experimental parameters as in Fig.~\ref{fig:boomerang}b. In contrast to the characteristic boomerang-like evolution when $\Delta=0$, each of the five experimental realizations of randomness for $\Delta=0.6$ exhibits a different behavior. The mean momentum over these disorder realizations demonstrates the destruction of the quantum boomerang effect, with a long-time average momentum significantly different from zero, and agrees reasonably well with the result of disorder-averaged simulations. Fig.~\ref{fig:theRandomness}b shows the evolution of the average energy for the same experiments: the destruction of dynamical localization by random kicking and the approach to classical linear-time energy growth are clearly visible in both experiment and theory, and are associated with the destruction of the quantum boomerang effect in Fig.~\ref{fig:theRandomness}a. To directly probe the dependence of the quantum boomerang on the disorder strength, we plot in Fig.~\ref{fig:theRandomness}c the disorder-averaged mean momentum after 25 kicks as a function of $\Delta$. Measured long-time $\langle p \rangle$ increases monotonically up to $\Delta\approx 0.5$ and agrees reasonably well with disorder-averaged numerical predictions. 
An interesting topic for further study is the relationship between the degree of destruction of the boomerang effect (measured, for example, by long-time average momentum) and the exponent characterizing energy growth~\cite{Ringot2000,HoogerlandPhaseNoise,RaizenAmplitudeNoise}.

In this work we have presented the first experimental evidence for the quantum boomerang effect in an Anderson-localized system. The measurements confirm theoretical predictions~\cite{PratPRA2019,TessieriPRA2021} and demonstrate that boomerang dynamics can occur from a variety of initial conditions including even a homogeneous density distribution. 
These experimental results, along with the analytical theory we present in the appendices, elucidate the crucial role of parity and time-reversal symmetry in determining the presence or absence of the quantum boomerang effect. 
We have probed the dependence of boomerang dynamics upon Anderson localization itself by measuring the effect of tunably random kicking sequences: results confirm that classical diffusion induced by stochasticity destroys both dynamical localization and the quantum boomerang effect. These experiments validate a new, powerful probe of the uniquely quantum-mechanical nature of localization applicable to a general class of disordered systems and suggest a variety of intriguing topics for future exploration.  These include boomerang phenomena in higher-dimensional systems,  different symmetry classes~\cite{SmilanskySymmetry92}, more exotic initial states, and the presence or absence of many-body boomerang effects in interacting systems~\cite{JanarekPRA2020} or even many-body localized states. Applications of our demonstration --- that adjusting the Floquet gauge can switch on or off the boomerang effect --- are an intriguing possibility. Beyond the context of the quantum kicked rotor, the investigation of boomerang-like phenomena in condensed matter systems and implications of these results for ultrafast electron dynamics in disordered solids represent unexplored frontiers.

\section*{Acknowledgments}
We thank Toshihiko Shimasaki, Esat Kondakci, and Yifei Bai for helpful conversations. D.W.\ acknowledges support from the Air Force Office
of Scientific Research (FA9550-20-1-0240), the Army Research Office (W911NF-20-1-0294),
the National Science Foundation (CAREER 1555313), and the Eddleman Center for Quantum Innovation, and from the NSF QLCI program through grant number OMA-2016245. T.M.\ acknowledges CNPq (grant no. 311079/2015-6). 
T.M.\ and F.S.\ are supported by the Serrapilheira Institute (grant number Serra-1812-27802). R.S.\ and E.N.-M.\ acknowledge support from the UCSB NSF Quantum Foundry through the Q-AMASEi program (Grant No.\ DMR-1906325). A.C. acknowledges support from the NSF Graduate Research Fellowship Program (grant DGE2040434).

\section*{Author Contributions}

R.S., J.T., H.M., E.N.-M.\, and E.S.\ performed the measurements. R.S., J.T., and H.M.\ analyzed the data. R.S., H.M., J.T., E.S., F.S., and A.C.\ performed numerical modeling. A.C., T.M., and F.S.\ performed the theoretical symmetry analysis with assistance from P.V. R.S., J.T., H.M., A.C., and D.W.\ drafted the manuscript with input from all authors. D.W.\ supervised the experiments. All authors contributed insights in discussions on the project. 

\appendix

\section{Deriving Conditions for the QKR Boomerang Effect}
\label{sec:alpha=1/2}
Here we show that the symmetry conditions on the initial state and Floquet gauge described in the main text indeed enable the boomerang effect in the QKR model. Explicitly, we consider the canonical QKR Hamiltonian considered in the main text (reproduced here and reverting to standard variable notations for clarity)
\begin{align}
    \mathcal{H}_{\alpha}(t) = \frac{p^2}{2} + K \cos x \sum_{n=-\infty}^\infty \delta(t-n-\alpha).
    \label{eq:H}
\end{align}
To be explicit, here we define $\alpha$ on the interval $\left[0,1\right]$, taking the cases $\alpha=0$ and 1 to be distinct from each other and consistent with the definition in the Floquet operator
\begin{align}
    U_{\alpha} = e^{-i (1-\alpha) p^2 /2 \kbar} e^{-i K \cos x / \kbar} e^{-i \alpha p^2/2 \kbar}.
    \label{eq:U}
\end{align}
That is, for $\alpha = 0 \,(1)$ the kicks occur just after (before) $t = n$. Here we continue to refer to symmetry operations only about fixed axes. $\mathcal{H}_{\alpha}$ is evidently parity inversion ($x \to -x$, $p \to -p$) symmetric at all times
\begin{align}
    \left[ \mathcal{P}, \mathcal{H}_{\alpha}(t) \right] = 0.
    \label{eq:parity}
\end{align}
The operator $\mathcal{P}$ obeys $\mathcal{P}^\dagger p \mathcal{P} = -p$. Only for $\alpha=1/2$ does the system also become time-reversal ($t \to -t$, $p \to -p$) symmetric. The complex conjugation operation $\mathcal{K}$ associated with time-reversal acts on the position space wave-function as $\bra{x} \mathcal{K} \ket{\psi} = \psi^*(x)$. For the $\alpha=1/2$ case, $\mathcal{H}_{\alpha}$ is additionally parity-time inversion ($t \to - t$, $x \to -x$) symmetric.

We  use the following intuitive definition for the QKR boomerang effect in momentum space. (i)~The boomerang should exhibit non-trivial average momentum dynamics at short times, in particular a departure from followed by a relaxation toward the initial value. (ii)~At late times, a true boomerang should have an average momentum equivalent to the initial momentum. Explicitly, we will show that (i) is only compatible with parity asymmetric states. Restricting ourselves to the case of 0 initial momentum, we then show that time-reversal symmetry of the quenched initial state and Floquet gauge guarantees (ii).

We consider solutions to the time-dependent Schrodinger equation $i \kbar \partial_t \ket{\psi(t)} = \mathcal{H}_{\alpha}(t) \ket{\psi(t)}$, where $\ket{\psi(t=0)}$ is the initial state quenched into the Hamiltonian. Our analysis will concern the behavior of $\langle p(t) \rangle = \bra{\psi(t)} p \ket{\psi(t)}$. First we consider the evolution of an arbitrary state $\ket{\psi(t=0)}$ and its parity inverted counterpart $\ket{\phi(t=0)} \equiv \mathcal{P} \ket{\psi(t=0)}$. Let $\langle p(t) \rangle$ and $\langle \tilde{p}(t) \rangle$ correspond to the time-evolved states $\ket{\psi(t)}$ and $\ket{\phi(t)}$ respectively. From parity symmetry (\ref{eq:parity}), we have $\ket{\phi(t)} = \mathcal{P} \ket{\psi(t)}$, and from this can conclude
\begin{align}
    \langle \tilde{p}(t) \rangle = \bra{\psi(t)} \mathcal{P}^{\dagger} p \mathcal{P} \ket{\psi(t)} = - \langle p(t) \rangle.
    \label{eq:p1=-p2}
\end{align}
This holds for any initial state and $\alpha$, as well as at all times $t$. From this, it immediately follows that a necessary condition for observing (i), and thus the boomerang, is that $\ket{\psi(t=0)} \neq \mathcal{P} \ket{\psi(t=0)}$; otherwise $\langle \tilde{p}(t) \rangle = \langle p(t) \rangle = 0$, guaranteeing that (ii) is satisfied but in a trivial way. Since all of the remaining analysis is also concerned with $\langle p(t) \rangle$, we hereafter assume that parity asymmetry is satisfied to derive non-trivial results.

Now we begin to investigate the $t \to \infty$ behavior concerning (ii). Considering the Floquet spectral decomposition of the initial state $\ket{\psi(t=0)} = \sum_m c_m \ket{\varphi_m^{\alpha}}$ with $U_{\alpha} \ket{\varphi_m^{\alpha}} = e^{-i \omega_m^{\alpha}} \ket{\varphi_m^{\alpha}}$, the long-time average momentum given by the Floquet diagonal ensemble is
\begin{align}
    \langle p(t \to \infty) \rangle = \sum_m \abs{c_m}^2 \bra{\varphi_m^{\alpha}} p \ket{\varphi_m^{\alpha}}. 
    \label{eq:FDE}
\end{align}
One can also consider the opposite limit $t\to -\infty$ by decomposing in the basis of $U_{\alpha}^{\dagger}$. Since the eigenstates of a unitary operator and its Hermitian-conjugate are identical (both are eigenstates of the same self-adjoint, time-independent Floquet Hamiltonian), the prediction of the Floquet diagonal ensemble is the same. This leads to the conclusion that
\begin{align}
    \langle p(t \to \infty) \rangle = \langle p(t \to -\infty) \rangle.
    \label{eq:p_inf=p_-inf}
\end{align}
This also holds for any initial state and $\alpha$.

Now finally we consider time-reversal symmetric states $\mathcal{K} \ket{\psi(t=0)} = \ket{\psi(t=0)}$. Under parity-time inversion, we then have the relation
\begin{align}
    \mathcal{P} \mathcal{K} \ket{\psi(t=0)} = \mathcal{P} \ket{\psi(t=0)} \equiv \ket{\phi(t=0)}.
    \label{eq:PKinit}
\end{align}
Now consider the full solution $\ket{\psi(t)}$ and $\langle p(t) \rangle$. For $\alpha=1/2$, the parity-time inversion symmetry of the Hamiltonian implies that $\mathcal{P} \mathcal{K} \ket{\psi(-t)}$ is another solution with a corresponding momentum evolution $\langle p(-t) \rangle$. From (\ref{eq:PKinit}), it is straightforward to see that this solution corresponds to the evolution of the parity-inverted initial state, i.e. $\ket{\phi(t)} = \mathcal{P} \mathcal{K} \ket{\psi(-t)}$; note that while a parity-time inverted state about a different time-axis can be defined for arbitrary $\alpha$, such a state will not coincide with $\ket{\phi(t)}$ and will consequently prevent the following conclusions. Defining $\langle \tilde{p}(t) \rangle$ in the same way as before, this result implies $\langle \tilde{p}(t) \rangle = \langle p(-t) \rangle$. Combining this with (\ref{eq:p1=-p2}), we thus have
\begin{align}
    \langle p(-t) \rangle = -\langle p(t) \rangle.
    \label{eq:p(t)_odd}
\end{align}
One can deduce this relation intuitively by noting that backward propagation under $U_{\alpha=1/2}$ for a time-reversal symmetric initial state simply corresponds to receiving kicks of the opposite sign. Finally, the combination of (\ref{eq:p_inf=p_-inf}) and (\ref{eq:p(t)_odd}) yields the final result
\begin{align}
    \langle p(t \to \infty) \rangle = 0.
    \label{eq:pinf=0}
\end{align}

We conclude this appendix by discussing the important role of quasi-momentum averaging and the applicability of this argument to the experiment. First, we point out that the experimentally relevant quasi-momentum averages are explicitly consistent with the presented argument for boomerang dynamics. In the primary experimental results presented, the initial state is an ensemble of ground-band Bloch states $\psi_{\beta}(x-x_0)$, where $\beta$ denotes quasi-momentum, and we reference the functional form $\psi_{\beta}(x)$ with respect to a cosine external potential. Generically, the entire Bloch ensemble will be parity asymmetric ($\psi(-x) \neq \psi(x)$) due to the non-zero shift $x_0$ of the peak density away from $x=0$, and thus condition (i) remains trivially satisfied. However, time-reversal symmetry demands a real spatial wave-function ($\psi^*(x) = \psi(x)$), and this is only true for $\beta=0$. The solution is to observe that the superposition $\psi_{\beta}(x) + \psi_{-\beta}(x)$ is real for arbitrary $\beta$, and thus the averaging can be performed so long as the quasi-momentum distribution is symmetric. Even though this argument is based on the coherent superposition of $\pm \beta$ states, the conclusion holds for thermal quasi-momentum distributions as the QKR dynamics are quasi-momentum conserving and the momentum operator does not mix quasi-momentum sectors. We note that a similar disorder symmetrization can also be applied to the Anderson model, where an explicit average over parity-conjugated disorder potential pairs $V(x)$ and $V(-x)$ ensures the validity of the above argument at the level of individual disorder realizations (switching $x$ for $p$ as well as the role of time and parity-time reversal). It is important to keep in mind, however, that such an explicit symmetrization of the disorder averaging is not required for observing the boomerang effect in the Anderson model; whether a boomerang effect exists for asymmetric quasi-momentum distributions in the QKR is an open question for future exploration.

Despite the fact that the argument presented in this section holds for individual pairs of quasi-momentum, we stress that quasi-momentum disorder averaging remains critical for practically observing boomerang dynamics. Our proof of (ii) is based on the assumption that the momentum expectation value relaxes in the long-time limit to the static diagonal ensemble prediction. However, individual quasi-momentum states will exhibit large momentum oscillations at all times and only relax in the sense of the time-averaged momentum. This is in contrast to our experimental results where the system has a finite-width quasi-momentum distribution (due more to finite size than to finite temperature), and subsequently the instantaneous momentum relaxes to 0, only exhibiting small fluctuations about this value for large times. We interpret the latter situation as more representative of a true boomerang effect.

\section{Time-Reversal Symmetry for Kicked State under Talbot Dynamics}
\label{sec:tablot/4_kickedstate}
Here we explicitly prove that the kicked initial state preparation indeed leads to a time-reversal symmetric state after free-evolution by $T_{\mathrm{Talbot}}/4$. For this analysis, we consider 0 quasi-momentum. Using the Jacobi-Anger expansion for the initial wavefunction $\psi(x) = e^{-i V \sin x}$ described in the main text, we can decompose the initial kicked state in a discrete momentum basis indexed by integers $l$ as
\begin{align}
    \ket{\psi(t=0)} = \sum_{l=-\infty}^{\infty} (-1)^l \mathcal{J}_l(V) \ket{l},
    \label{eq:kickedstate}
\end{align}
where $\mathcal{J}_l$ is the $l^\mathrm{th}$ Bessel function of the first kind. The quarter-Talbot period free evolution is described by $e^{-i p^2/2 \kbar}$ with $\kbar=\pi$ and $p = \kbar l$. Noting that $(-1)^l = (-1)^{l^2}$ we have
\begin{align}
    \ket{\psi(t = T_{\mathrm{Talbot}}/4T)} = \sum_{l=-\infty}^{\infty} i^{l^2} \mathcal{J}_l(V) \ket{l}.
\end{align}
The condition on the momentum coefficients for a time-reversal symmetric or real position-space wavefunction is $c^*_{-l} = c_l$, where $c_l = \bra{l} \ket{\psi}$. Using $\mathcal{J}_{-l}(V) = (-1)^l \mathcal{J}_{l}(V)$, we straightforwardly confirm this condition
\begin{align}
    c^*_{-l} = (-i)^{(-l)^2} \mathcal{J}_{-l}(V) = i^{l^2} \mathcal{J}_l(V) = c_l. 
\end{align}
Taken with the proof in appendix \ref{sec:alpha=1/2}, this supports the claim that the kicked state should exhibit a boomerang effect for $\alpha = 1/2 + T_{\mathrm{Talbot}}/4T$. We note that this argument is only exact for 0 quasi-momentum, and the experimental kicked initial state data relies on a narrow quasi-momentum distribution about 0 for clear observation of the boomerang effect; however in considering corrections to this argument, one should keep in mind that the initial state (\ref{eq:kickedstate}) considered in this analysis is already only an approximate description of the experimental reality.

\section{Floquet Gauge Symmetry Properties of Late-Time Momentum}
\label{sec:palpha}

Here we prove that the infinite-time momentum is either an even or odd function of $\alpha$ about 1/2 for states with the given symmetry properties explored in the experiment. First we consider initial states with time-reversal symmetry such that
\begin{align}
    \mathcal{K} \ket{\psi(t=0)} = \ket{\psi(t=0)}.
    \label{eq:trsstate}
\end{align}
We note the relation
\begin{align}
    \mathcal{K}^{\dagger} U_{\alpha} \mathcal{K} = e^{i (1-\alpha) p^2/2 \kbar} e^{i K \cos x / \kbar} e^{i \alpha p^2/2\kbar} = U_{1-\alpha}^{\dagger}.
    \label{eq:conjU}
\end{align}
For compactness, we will henceforth denote the state after $n$~kicks in the gauge $\alpha$ as $\ket{\psi_n}_{\alpha} \equiv U_{\alpha}^n \ket{\psi(t=0)}$ ($\ket{\psi_0} = \ket{\psi(t=0)}$) and use similar notation for expectation values. The momentum evolution in the gauge $1-\alpha$ is then given by
\begin{align}
    \langle p_n \rangle_{1-\alpha} = \bra{\psi_0} U_{1-\alpha}^{n \dagger} \, p \, U_{1-\alpha}^n\ket{\psi_0}.
\end{align}
Using (\ref{eq:trsstate}), (\ref{eq:conjU}) and $\mathcal{K}^{\dagger}\, p \,\mathcal{K} = -p$, we find
\begin{align}
    \label{eq:p(n)_1-alpha=-p(-n)_alpha_step1}
    \langle p_n \rangle_{1-\alpha} &= \bra{\psi_0} \mathcal{K}^{\dagger} U_{1-\alpha}^{n \dagger} \mathcal{K} \mathcal{K}^{\dagger} p \mathcal{K} \mathcal{K}^{\dagger} U_{1-\alpha}^n \mathcal{K}\ket{\psi_0} \\
    &= - \bra{\psi_0} U_{\alpha}^n \, p \, U_{\alpha}^{n \dagger} \ket{\psi_0} = - \langle p_{-n} \rangle_{\alpha}.
\end{align}
That is, forward quench dynamics in the gauge $1-\alpha$ possess an opposite momentum evolution to backward evolution under the gauge $\alpha$. Note that this is essentially a generalization of equation (\ref{eq:p(t)_odd}) to arbitrary $\alpha$; evaluating this property at $\alpha=1/2$ in conjuction with (\ref{eq:p_inf=p_-inf}) serves as an alternative proof that the late-time zero momentum feature of the boomerang effect only depends on simultaneous time-reversal symmetry of the Floquet gauge and quenched state. Leaving $\alpha$ arbitrary, taking the limit $n \to \infty$, and applying (\ref{eq:p_inf=p_-inf}), we derive the odd function result for time-reversal symmetric states
\begin{align}
    \langle p_{\infty} \rangle_{1-\alpha} = - \langle p_{\infty} \rangle_{\alpha}.
    \label{eq:palpha_odd}
\end{align}

Next we consider a state with the symmetry
\begin{align}
    \mathcal{P} \mathcal{K} \ket{\psi(0)} = \ket{\psi(0)}.
\end{align}
Replacing $\mathcal{K} \to \mathcal{P} \mathcal{K}$ in (\ref{eq:p(n)_1-alpha=-p(-n)_alpha_step1}), making use of the parity symmetry of the evolution (\ref{eq:parity}), and noting that $\mathcal{K}^{\dagger} \mathcal{P}^\dagger p \mathcal{P} \mathcal{K} = p$, we instead find
\begin{align}
    \langle p_n \rangle_{1-\alpha} = \langle p_{-n} \rangle_{1-\alpha}.
\end{align}
Taking the infinite-time limit in the same way as before, we arrive at the desired even function result
\begin{align}
    \langle p_{\infty} \rangle_{1-\alpha} = \langle p_{\infty} \rangle_{\alpha}.
    \label{eq:palpha_even}
\end{align}
Here we suggest an intuitive understanding for the difference between the two cases. Whereas the backward evolution in $\alpha$ for a time-reversal symmetric state only differs from the forward evolution in $1-\alpha$ by the sign of the kicks, which leads to opposite sign momentum impulses and subsequently (\ref{eq:palpha_odd}), the backward free-evolution in $\alpha$ of the parity-time reversal symmetric state produces an additional parity inversion with respect to the forward free-evolution in $1-\alpha$ (for instance, see Fig.~\ref{fig:kickprep}c), cancelling the kick inversion to yield (\ref{eq:palpha_even}).

\section{TDSE and Diagonal Ensemble Numerics}
\label{sec:numerics}
In all TDSE calculations discussed in this work we compute the dynamics of individual Bloch states and average over a Gaussian quasimomentum distribution of standard deviation $0.1\,k_{\mathrm{L}}$, which both models our finite-width BEC and realizes the disorder average in standard Anderson model calculations.~\footnote{According to the adiabatic theorem, the BEC wavefunction after the lattice load state-preparation is a superposition of Bloch states $e^{i \beta x}u_{\beta}(x)$, where $\beta$ denotes quasimomentum, weighted by the initial Gaussian momentum wavefunction ($\psi(p)$ before the load becomes $\psi(\beta)$) and with a $\beta$-dependent phase. Due to the periodicity of $u_{\beta}(x)$, it can be shown that the $\beta$-dependent phase does not affect the time-evolution of $\langle p \rangle$ and $\langle p^2/2M \rangle$.}. The simulation mesh contains $2^{10}$ points and the kicks are modeled as Gaussian pulses with $\sigma$ of $125\,\mathrm{\upmu s}$, calculated by fitting the pulses experimentally measured on an oscilloscope. During the free expansion stages of the experiment we numerically evolve the wavefunction in the momentum basis, while during the kicks we utilize a split-step method. We evolve the wavefunction with the kinetic energy operator in the momentum basis for half a time step, Fourier transform the wavefunction into the position basis and evolve with the kick operator (which is diagonal in the position basis) for a full time step, then finally Fourier transform back into the momentum basis and apply the kinetic energy operator for another half time step. This is repeated for all time steps over the duration of the pulse. We average over a uniform distribution of kicking lattice depths with a 10\% width to account for the atoms radially sampling the measured Gaussian cross section of the lattice, as well as drift in the kicking laser power. We suspect that a portion of the remaining discrepancy between theory and experiment results from slight differences in initial conditions arising from imperfect lattice loading, along with potential undersampling of data points. 

In Figs. 2e and 3b, we compute the Floquet diagonal ensemble predictions by diagonalizing the ideal QKR Floquet operator \ref{eq:U}. The diagonal ensemble momentum \ref{eq:FDE} is computed for ideal states which should exhibit the Floquet gauge symmetries discussed in appendix C. For comparison with the load-preparation data, we use a Gaussian Bloch state ensemble, while for comparison with the kick-preparation data, we use $\psi(x) = e^{-i V \sin(2k_{\mathrm{L}}x)}$ with $V$ the effective preparation kick impulse. Quasimomentum averaging is accomplished via the same method used in the TDSE numerics.

\providecommand{\noopsort}[1]{}\providecommand{\singleletter}[1]{#1}%

\end{document}